\documentclass[sigconf]{acmart}
\usepackage{multirow}

\AtBeginDocument{%
  \providecommand\BibTeX{{%
    \normalfont B\kern-0.5em{\scshape i\kern-0.25em b}\kern-0.8em\TeX}}}

\setcopyright{acmlicensed}
\begin{document}

%%
%% The "title" command has an optional parameter,
%% allowing the author to define a "short title" to be used in page headers.
%%
%% The "author" command and its associated commands are used to define
%% the authors and their affiliations.
%% Of note is the shared affiliation of the first two authors, and the
%% "authornote" and "authornotemark" commands
%% used to denote shared contribution to the research.
\title{Domain-adapted Learning and Interpretability: \\
DRL for Gas Trading}

%%
%% The "author" command and its associated commands are used to define
%% the authors and their affiliations.
%% Of note is the shared affiliation of the first two authors, and the
%% "authornote" and "authornotemark" commands
%% used to denote shared contribution to the research.
\author{Yuanrong Wang}
%\email{yuanrong.wang@cs.ucl.ac.uk}
\affiliation{%
  \institution{University College London}
  \institution{Shell Global Solutions International}
  \country{UK}
}

\author{Yinsen Miao}
\authornote{This work was performed while at Shell. Currently with Fidelity Investments, USA}
%\email{yinsen.miao@fmr.com}
\affiliation{%
  \institution{Shell Global Solutions International}
  \country{USA}
}

\author{Alexander CY Wong}
\authornote{This work was performed while at Shell. Currently with Mizar.}
%\email{alexwong\_92@hotmail.com}
\affiliation{%
  \institution{Shell Global Solutions International}
  \country{The Netherlands}
}

\author{Nikita P Granger}
%\email{Nikita.Granger@shell.com}
\affiliation{%
  \institution{Shell Energy North America}
  \country{USA}
}

\author{Christian Michler}
%\email{C.Michler@shell.com}
\authornote{Corresponding author.}
\affiliation{%
  \institution{Shell Global Solutions International}
  \country{The Netherlands}
}

%%
%% By default, the full list of authors will be used in the page
%% headers. Often, this list is too long, and will overlap
%% other information printed in the page headers. This command allows
%% the author to define a more concise list
%% of authors' names for this purpose.
\renewcommand{\shortauthors}{Y. Wang et al.}

%%
%% The abstract is a short summary of the work to be presented in the
%% article.
\begin{abstract}
Deep Reinforcement Learning (Deep RL) has been investigated for several applications in finance and trading, with a particular emphasis on the design of complex models and simulation environments. However, much of the literature has neglected the expertise of the trading field, resulting in limited knowledge that could improve agent performance and the interpretability of the methodology. This paper addresses this issue by providing insights into the feature selection process and the walk-forward training scheme used in a practical application of Deep RL to trade front-month natural gas futures contracts. Additionally, the analysis discusses model explainability with trading frequency and risk measures. Our implementation outperforms rule-based benchmarks and results reported in the literature, as demonstrated by a reported Sharpe Ratio. We also propose a simple yet effective ensemble learning scheme to stabilize agent actions, leading to enhanced model stability and robustness, as well as lower turnover and transaction costs.
\end{abstract}

%%
%% The code below is generated by the tool at http://dl.acm.org/ccs.cfm.
%% Please copy and paste the code instead of the example below.
%%

%%
%% Keywords. The author(s) should pick words that accurately describe
%% the work being presented. Separate the keywords with commas.

%% A "teaser" image appears between the author and affiliation
%% information and the body of the document, and typically spans the
%% page.

%\received{20 February 2007}
%\received[revised]{12 March 2009}
%\received[accepted]{5 June 2009}

%%
%% This command processes the author and affiliation and title
%% information and builds the first part of the formatted document.
\maketitle

\section{Introduction}

The profitability of a trading strategy hinges on a good timing of entering and exiting a position. Researchers, traders and quants have been exploiting fundamental and technical analysis to analyze the market with the aim of predicting future market movements ever since. However, among the ever increasing complexity and the notoriously low signal-to-noise ratio of financial markets, many expert-designed rule-based strategies fail to cope with changing market conditions. Recent advancements in Artificial Intelligence and data science start to bring a competitive edge to trading by learning the market dynamics. With the vast amount of historical data, models' non-linear learning and inference ability extract patterns in time series and exploit volatility and directional signals for position taking. 

Natural gas is one of the most liquid commodity markets. Energy futures and options contracts are usually traded on derivative marketplaces such as the Chicago Mercantile Exchange (CME) or over-the-counter. Given the high volatility and similar market dynamics between gas and equity markets, the classic technical indicators are considered to still provide analytical signals for the underlying gas price. However, as commodities are influenced more by supply and demand, the fundamental analysis is very different from stocks that are mostly correlated to the valuation of underlying firms. Key contributing factors to the gas price movements are often macro economic, including regional storage and production, global demand, alternative power production as well as weather. These differences prevent many stock traders to enter the energy market. Nevertheless, some trading algorithms have shown to be transferable, see e.g.~\citep{Kanamura2009APM, Lubnau2015TradingOM,Husby2019CanTT}.

Broadly speaking, quantitative models in commodity trading focus on the following aspects: Event-driven traders react to system events and failures. Similar to how corporate news influences the price of stocks, a failure in a transmission system or a generator \citep{Hafez2018MachineLA,Ma2019TheDO} and sometimes in a policy \citep{Rosa2013TheHR} can greatly impact the price of energy. Then, technical analysis and fundamental analysis are two main methodologies in the market, where strategies learn from past data for future price forecasting. An early success story of Moshiri \citep{Moshiri2006ForecastingNC} employed a non-linear Artificial Neural Network model to predict crude oil futures which outperformed the traditional econometric-based models. Since then, ensemble learning methods by Yu and Jammazi \citep{Yu2008ForecastingCO, Jammazi2012CrudeOP} combined Machine Learning (ML) and econometric models by Godarzi and Zhang \citep{Godarzi2014PredictingOP,Zhang2015ANH}, and also deep learning models by Zhao, Tang, Zhao and Safari \citep{Zhao2017ADL,Tang2015ANC,Zhao2018ANM,Safari2018OilPF} have been proposed and designed. More recently with the advancement in hardware and execution speed, both aggressive and passive high frequency trading algorithms have been applied in commodity trading, see \citep{Miller2016HighFT}. As the order book structure in natural gas futures on CME and equities on stock exchanges are similar, momentum ignition, order anticipation and arbitrage trading have been explored, see, e.g., Fishe \citep{Fishe2018HighFT}. 

Despite the abundance of data and learning ability of ML models, many issues arise from the noisy time-series data, especially in finance, and biased supervised labelling in the trading environment. An extremely low signal-to-noise ratio in financial time-series is an inherent impediment despite considerable efforts in time-series filtering by means of, e.g., information filtering networks \citep{Tumminello2005ATF,Massara2017NetworkFF,Massara2019LearningCF,Wang2022SparsificationAF}. Moreover, the general two-step approach in constructing a trading strategy is ill-posed, see~\citep{Moody1998PerformanceFA}: Firstly, a supervised model forecasts asset price changes with a defined investment horizon. Then, the forecasts are fed to a strategy module to generate actual trading strategies. In the first step, the supervised models are normally labelled by future prices with a defined investment horizon. This approach limits generalization of the model. In addition, besides the predictions, other features, e.g., liquidity, market micro-structure, are usually not included in the second step. This ideal simplification of market dynamics fails to explicitly address the interaction between a trading strategy and the real market in the form of market impact during execution. 

Reinforcement Learning (RL) agents learn policies (strategies) directly by optimising a numerical reward signal by interacting with a virtual market environment that is usually derived from historical market data. This has been shown to prevent many of the aforementioned issues in supervised ML by Sutton \citep{Sutton2005ReinforcementLA}. Instead of back-propagating the loss between labelled ground truth and the prediction, a simulation environment is built based on market data for Deep RL agents to explore and exploit. Next, actions are evaluated based on the reward from the interaction with the market environment, and a policy (trading strategy) is learned through iterative interactions. In 2013, Mnih\citep{Mnih2013PlayingAW} published the seminal work of Deep Q-learning Network (DQN), which marks the transition from Reinforcement Learning to Deep Reinforcement Learning. Since then, more algorithms have been proposed, e.g, PPO~\citep{Schulman2017ProximalPO}, and DDPG~\citep{Lillicrap2016ContinuousCW}. Early literature has already attempted to apply traditional RL in the financial market, like stock pricing and selection \citep{Lee2007AMA,Eilers2014IntelligentTO,Tan2011StockTW}, optimal trade execution~\citep{Bertsimas1998OptimalCO, Kakade2004CompetitiveAF,Nevmyvaka2006ReinforcementLF}, high frequency trading~\citep{Sherstov2004ThreeAS,Usmani2015AnII} and portfolio trading~\citep{Neuneier1997EnhancingQF,Jin2016PortfolioMU,Gudimella2017DeepRL}. In the last few years, Spooner \citep{Spooner2018MarketMV} has used several Time-Difference methods for market making. \citep{Xiong2018PracticalDR} applied DDPG for stock trading. RoaVicens \citep{RoaVicens2019TowardsIR} has embarked on simulating an order book environment in the presence of competitive agents. Yang \citep{Yang2020DeepRL} researched ensemble methods between different Deep RL agents. Other recent Deep RL research in financial applications has been summarized by Hambly \citep{Hambly2021RecentAI}. Besides the academic community, JP Morgan in 2018 has released a working paper \citep{Bacoyannis2018IdiosyncrasiesAC} introducing their RL based limit order engine highlighting the potential for applications in trading.

In 2019, Zhang \citep{Zhang2019DeepRL} compared DQN, PG, A2C algorithms for equity, FX, fixed income and commodity trading as the current state-of-the-art benchmark. Our work builds on the structure of the DQN applied to commodity trading~\citep{Zhang2019DeepRL} and combines it with the implementation from Udacity \footnote{https://github.com/udacity/deep-reinforcement-learning}. For performance improvement and model explainability, we further incorporate a game-theoretical SHAP (SHapley Additive exPlanations) plot from Lundberg\citep{Lundberg2017AUA} for feature interpretation and selection. Furthermore, to address the high compute cost of training Deep RL models discussed in Gudimella \citep{Gudimella2017DeepRL}, we leverage the Microsoft Bons.ai platform~\citep{Bons.ai2017} with its containerized simulation environments, high parallelism and automated hyper-parameter tuning which accelerates model training and testing. Yang \citep{Yang2020DeepRL} and Carta\citep{Carta2021AMA} emphasized that the same model with different initializations can lead to different outcomes due to the long path-dependency especially for problems in trading. To prevent bias from back-testing and enhance robustness of the model,  Borokova \citep{Borovkova2019AnEO} present an online ensemble learning method for LSTM in high-frequency equity price prediction, where models are weighted by their recent performance. Inspired by their framework, we propose and design an ensemble learning scheme, where agents are trained in parallel before filtering by training performance, and then averaged by threshold voting. The implementation details are presented in Section \ref{sec:Methodology}, and results are presented in Section \ref{sec:Results} and discussed in Section~\ref{sec:ME} for interoperability and Section~\ref{sec:Discussion} for performance.

\section{Problem description} \label{sec:Problem description}

\subsection{Data} 
We trade the front month NBP UK Natural Gas Futures contract. The future contracts are for physical delivery through the transfer of rights in respect of Natural Gas at the National Balancing Point (NBP) Virtual Trading Point, operated by National Grid, the transmission system operator in the UK. Delivery is made equally each day throughout the delivery period. The NBP virtual trading point acts as a central exchange. The price during a trading day can be characterized by a so-called candle of low, high, open and close price \citep{Morris1995CandlestickCE}. Alongside the trading volume of the day, we shall refer to these five features as price features. Other than the price features, technical analysis is employed to construct technical features. In this experiment, we compute MACD, Price RSI, Volume RSI, PCA 1st component and volatility adjusted returns of 1, 2 and 3 months as additional features \citep{Diebold2010TheKT}. Furthermore, to investigate the effect of natural gas fundamental features, we also include demand data (industrial, gas to power and residential demand), production data from UK production fields, LNG data in all 3 UK terminals, data of pipeline imports from Norway, Netherlands, Belgium and Ireland as well as storage data in all active facilities. A full list of features is shown in Table \ref{fig:feature table}.

\begin{table*}[h]

\begin{center}\begin{tabular}{|c|c|}
\hline 
{\bf Technical Features} & {\bf Fundamental Features} \\\hline 

\multirow{1}{*} {\bf Candlestick:} & {\bf Demand Data} \\   OHLCV & (Industrial, Gas to Power, Residential)  \\ \hline

\multirow{1}{*} {\bf Difference:} & {\bf Production Data} \\    High - Low price of the day, & (UK production fields) \\ High - Open price of the day, & {}  \\ Close - previous Close price of the day & {} \\ \hline

\multirow{1}{*} {\bf Technical Indicators:} & {\bf LNG Data} \\    MACD & (All 3 UK LNG terminals) \\ Price RSI, Volume RSI & {} \\ PCA $1^{st}$ component & {} \\ \hline

\multirow{1}{*} {\bf Returns:} & {\bf Pipeline Data} \\    Volatility adjusted 1-month & (Imports from NO, NL, BE, IR) \\ Volatility adjusted 2-month & {} \\ Volatility adjusted 3-month & {} \\ \hline

\multirow{1}{*}  & {\bf Storage Data} \\   {}  & (All active facilities)  \\ \hline

\end{tabular} 

\vspace{0.5pt}

	\caption[Feature Table]{Table outlines the technical and fundamental features used in the experiment. Technical features include candlestick features and their derived differences, technical indicators and volatility adjusted returns. Fundamental features include demand, production, transportation and storage data as well as liquefied natural gas (LNG) terminal data in the UK.}
	\label{fig:feature table}
\end{center}
\end{table*}

\section{Deep RL Methodology}\label{sec:Methodology}
\subsection{Set up of virtual market simulation environment} 
In Reinforcement Learning, agents learn policies from interacting with a simulation environment. Hence the environment needs to be as realistic as possible. A {\it complete market} approximates the real market where friction, market impact from orders, transaction cost and asset liquidity exist. However, such a highly interactive environment is complex to model and computationally demanding to simulate. Hence, a simplified {\it incomplete market} is customarily used where only transaction cost is considered. Here, we shall adopt the latter approach and consider transaction costs of $0.1\mathrm{p/therm}$. In this environment, agents observe the price, technical and fundamental features at each time step with a look-back window to detect trends. The agent then outputs an action for this time step, which is evaluated according to a reward function that guides and incentivizes the learning.

We limit this experiment to a daily trading frequency, and add a three-day look-back window to the aforementioned features in table \ref{fig:feature table}, which comprise our {\bf state space}. Therefore, the transition in the observational space moves forward the time-stamp by a day. We limit the size of the look-back window so that agents consider the most recent information, and the specific choice of the 3-day look-back window follows from a trade-off between model complexity and training efficiency. That is, each day an agent perceives all the information from the present day and the three previous days\footnote{Note that for the day under consideration the agent can only see the open price of the day, but not the close, high or low price of the day to prevent leaking future data that would only be known after trading closes on that day.}. As the trading environment is set up for daily frequency, the agent outputs the action for today based on the learned policy. For simplicity, the {\bf action space} is taken to be discrete as buy or sell the maximum amount, or hold, $A_t \in \{0, 1, -1\}$. To avoid overfitting the data in the replay buffer, a $1\%$ Gaussian noise is added to each observation.

Learning optimal decisions is guided by the reward function as a result of the agent's interaction with the environment through actions~$A_t$, and the {\bf reward}~$r_{i,t}$ in form of raw P\&L. In addition, for simulation purposes, we fixed the transaction costs, $tc = 0.1\mathrm{p/therm}$ based on the bid-ask spread. Thus, the total transaction cost per position change is $tc |A_t - A_{t-1}|$. We employ reward shaping as follows: The immediate reward after an action is taken as the P\&L, $\hat{r}_{i,t}$ to guide the agent, and the performance of each episode is assessed at the end of an episode by the annualized Sharpe Ratio~$SR_i$, where subscript~$t$ indicates the time-step in episode~$i$. 

\begin{equation}
	\begin{aligned} 
		\hat{r}_{i,t}=A_{t-1} r_{i,t} -& tc |A_t - A_{t-1}|\\
		%    \hat{r}_{i,t}=A_{t-1} \times r_{i,t} -& tc \times |A_t - A_{t-1}|\\
		\hat{\bar{r}}_i=\frac{1}{n}\sum_{t=1}^n \hat{r}_{i,t} \; ; \;\;\;\hat{\sigma}_i&=\sum_{t=1}^n \sqrt{\frac{1}{n}{|\hat{r}_{i,t}-\hat{\bar{r}}_i|^2 }} \\
		\hat{SR}_i = &\sqrt{252} \frac{\hat{\bar{r}}_i}{\hat{\sigma}_i}.
	\end{aligned}
\end{equation}

\subsection{Implementation}

The models are implemented separately in two environments. The Microsoft Bons.ai platform\footnote{https://www.microsoft.com/en-us/ai/autonomous-systems-project-bonsai?activetab=pivot\%3aprimaryr7} is an automated platform for Deep RL, where algorithms are standarized and hyper-parameter tuning is performed automatically. We leverage APEX DQN \citep{Horgan2018DistributedPE} as a discrete-action-spaces agent in our experiment to analyze different walk-forward training schemes. 

On the other hand, we have built a version of the in-house code with DQN algorithms based upon the Udacity standard framework \footnote{https://github.com/udacity/deep-reinforcement-learning/tree/master/dqn}, which serves as a verification and complement to the Bons.ai platform. Its main edge is the fine-grained control over the low-level neural network and simulation architectures, which benefits the design of our ensemble learning method referred to as {\it Filtered-Thresholding} detailed in Section \ref{sec:ensemble learning for vb}. To better match the high computation speed in Bons.ai, a parallel training scheme has been designed not only in the local version but also in a high-performance-computing version.  The in-house version has been used in our experiments for feature analysis, ensemble learning and model interpretation.

Under the standardized framework of Bons.ai, we specify the distributed actors to use a 2-layer Multi-layer Perceptron (MLP) of 64, 32 units with Rectifying Linear Units (ReLU) as the activation function. Then, to make a fair comparison, the in-house DQN agents also employ the same 2-layer MLP with ReLU. The detailed optimal hyper-parameter values of each agent during our experiments are listed below in Table \ref{tab:hp} for reference, where $\alpha$ is the learning rate, $\gamma$ is the discounting factor and $\tau$ is the interpolation parameter for soft updating.

\begin{table}[h]
\setlength\tabcolsep{1.5pt}
\begin{center}\begin{tabular}{|c|c|c|c|c|c|c|}
\hline 
{\bf Agent} & {\bf $\alpha$} & {\bf Optimiser} & {\bf Batch Size} & {\bf $\gamma$} & {\bf $\tau$} & {\bf Buffer Size}\\\hline 
APEX DQN & $ 10^{-5}$ & Adam & 64 & 0.6 & $ 10^{-3}$ & $10^{5}$\\\hline 
DQN & $ 10^{-5}$ & Adam & 64 & 0.6 & $ 10^{-3}$ & $10^{5}$\\\hline 
 
\end{tabular} 

\vspace{0.5pt}

\caption[HyperParameter]{Optimal Hyper-parameter values.
}
	\label{tab:hp}
\end{center}
\end{table}

\subsection{Ensemble Learning for Virtual Book} \label{sec:ensemble learning for vb}
During our extensive experiments, the initialization of different agents has been found to be an influential factor for the learning trajectory. An exact setup could even result in one agent learning successfully until converge while another agent being stuck in a local minimum. Moreover, as the underlying algorithms are model-free Deep RL, even all ``successfully'' learned agents will behave differently in terms of their underlying trading logic, e.g., one agent may be relying more on momentum strategy while the other agent may tend more towards mean-reversion trades. The strategy preference could be inferred from the difference in holdings across different agents in the same setup, which has been observed in an analysis to holding positions in Section \ref{sec: REL}.

\begin{figure}[h!]
	\centering
	\includegraphics[scale=0.32]{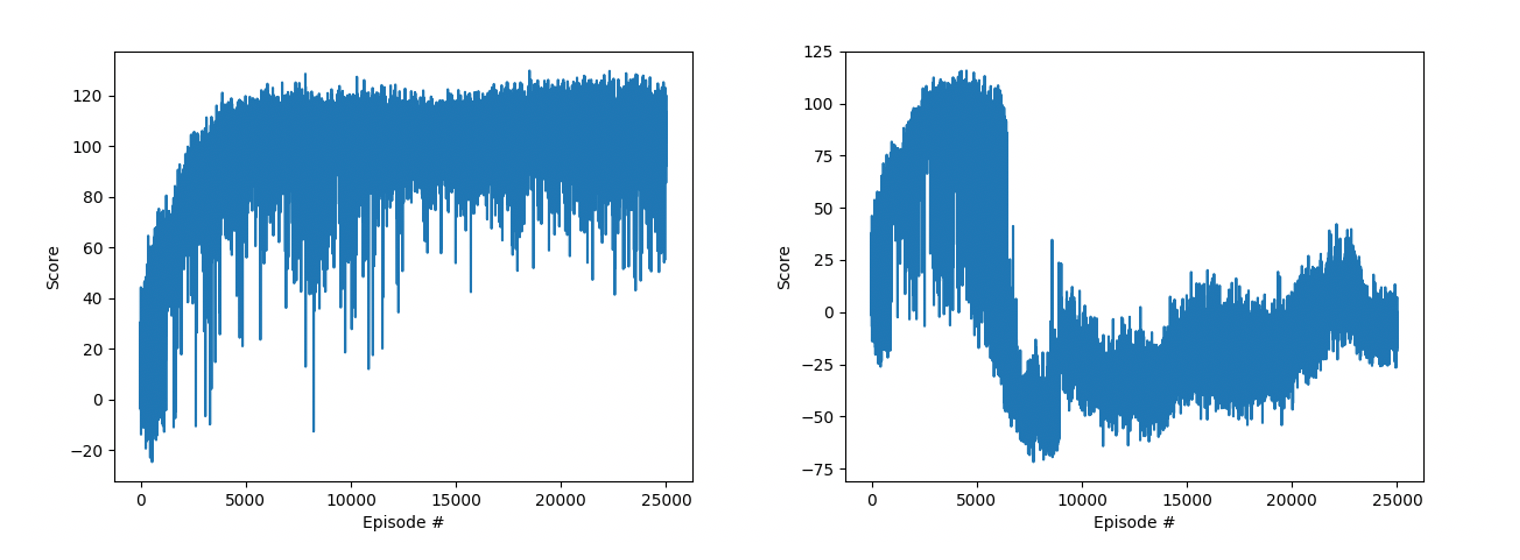}
	\vspace{0.5pt}
	\caption[Ensemble Learning Filtering]{Examples of a training curve for two instances of a Deep RL agent trained over all episodes during 2018 and 2019, showing the convergence of the trained agent~{\it{(left)}} and another agent getting stuck in a local minimum~{\it{(right)}}. }
	\label{fig:EL_filtering}
\end{figure}

To moderate the impact from sub-optimal agents and difference between underlying strategies, we propose a simple but effective ensemble learning method, which we refer to as {\it Filtered-Thresholding}. Firstly, $N$ number of agents are trained in parallel. Then, based on the trainings curve, we exclude seemingly sub-optimal agents in a ``Filtering'' step. After filtering out bad candidates, we ensemble the decision-making process between converged agents. Multiple ensemble methods have been considered, and the thresholding method was chosen. Each agent at each time-step has three choices / positions that it can take, namely Buy, Sell and Hold. If more than $p\%$ of agents agree on an action, then the ensemble agent executes the decision, otherwise it remains unchanged. In our experiment, $p\%=50\%$ is set for simplicity. This ``Thresholding'' step avoids large turnover in transaction costs caused by extensively moving in and out of positions, it reconciles any dissimilarities and integrates the advantages of the underlying agents and their respective trading logic.

An example of a 'successful' and a sub-optimal training curve is presented in Figure \ref{fig:EL_filtering}. In the left sub-plot, the score improves gradually until saturation and then remains more or less constant on average. However, in the right sub-plot, after an uptrend, the score drops into a local minimum and gets stuck. These training curves illustrate the criteria for the primitive selection process in ensemble learning. Automated systematic selection criteria can be obtained by combining the rolling average of the learning curve and monotone convergence theorem given in \citep{Bibby1974AXIOMATISATIONSOT}.

\section{Results}
\label{sec:Results}

\subsection{Walk Forward Training Schemes}

In any time-series machine learning problem, the training period is a crucial parameter that impacts the behaviour and performance of the algorithm. Besides computational cost, a long training period recognizes longer-term tendencies, while a short training period captures more local temporal patterns. This is especially true for Deep RL. Although prioritised experience replay attempts to emphasize the most recent pattern, the low signal-to-noise ratio inherent in financial time series still poses a challenge. Therefore, we compare two walk-forward training schemes, anchored and sliding-window approaches.

From a 12-year dataset of 2009 to 2020, the two approaches both start with a minimum 4-year training period and evaluate performance in the subsequent year. Then for the anchored approach, after each evaluation, the year that has just been used as off sample test set is added to the training period for the next walk-forward step, and the size of the training window grows over time. On the other hand, the sliding-window approach keeps the training window length remained fixed at 4 years while incrementing the start and end year of training.

\begin{table}[h]
\setlength\tabcolsep{2pt}

\begin{center}\begin{tabular}{|c|c|c|c|c|}
\hline 
{\bf Agent} & {\bf Scheme} & {\bf Avg. SR.} & {\bf MDD (\%)} & {\bf Cum P\&L (M\textsterling)} \\\hline 
APEX DQN & Anchor & 1.27 & 15.0& 44.95 \\\hline 
APEX DQN & Sliding & 1.32 & 22.3& 48.32\\\hline 
 
\end{tabular} 

\vspace{0.5pt}

\caption[Window Apporach]{Result table comparing the performance of anchored and sliding-window training schemes. The average Sharpe Ratio, maximum draw-down and cumulative P\&L are reported for APEX DQN trained on the Bons.ai platform.
}
	\label{fig:BonsaiSummery}
\end{center}
\end{table}

The performance summary of the two walk-forward schemes with APEX DQN are shown in Table \ref{fig:BonsaiSummery}. The table compares the performance of the two training schemes in terms of cumulative P\&L, average Sharpe Ratio and maximum draw-down. The APEX DQN sliding-window has an average Sharpe Ratio of 1.32 and a cumulative P\&L of 48.32 million GBP, compared to a lower performance of the anchored window with a Sharpe Ratio of 1.27 and a cumulative P\&L of 44.95 million GBP. The above summary statistics suggest a stronger performance of the sliding-window approach. This advantage mainly comes from excluding far away history as information to noise ratio decreases as further back we look in time. Moreover, the current model does not emphasize the most recent information, therefore a longer time horizon averages out the immediate influences. However, a shorter window of training history also leads to high volatility in inference, which is demonstrated by the 22.3\% maximum draw-down in the sliding-window approach compared to 15.0\% in the anchored window approach. Therefore, both schemes are valid training approaches, but the sliding window approach has a shorter training window and accordingly a faster convergence.   

\begin{table}[h]
\setlength\tabcolsep{2pt}

\begin{center}\begin{tabular}{|c|c|c|c|}
\hline 
{\bf Agent}  & {\bf Avg. SR.} & {\bf MDD (\%)} & {\bf Cum P\&L (M\textsterling)} \\\hline 
RL Selector MACD/BB  & 0.55 & 79 & 8.81 \\\hline 
MACD  & 0.18 & 174 & -1.17 \\\hline 
BB  & 0.00 & 210 & -4.25 \\\hline 
BUY \& Hold  & -0.18 & 252 & -6.39 \\\hline 
 
\end{tabular} 

\vspace{0.5pt}

	\caption[Rule Based Agents]{Result table for rule-based traditional trading methods served as baseline benchmarks. The average Sharpe Ratio, maximum drawdown and cumulative P\&L are reported for simple Buy\&Hold, MACD, BB, and naive RL selector between MACD and BB.}
	\label{fig:RuleBased}
\end{center}
\end{table}

To illustrate the superior performance of Deep RL agents, we present the three classic rule-based trading strategies as benchmarks and an RL selector with the identical evaluation metrics in Table \ref{fig:RuleBased}. The three benchmarks are naive buy and hold of the underlying asset as well as trading based on the 2 separate technical indicators, MACD and Bollinger Band. The RL selector is a naive RL agent to predict the most suitable indicator to follow based on the simulated market environment. 

The negative P\&L in all three rule-based strategies is unfortunate, but unavoidable. These results advocate the ever-increasing complicated trading environment where the once pioneered strategies all become insufficient. The added RL selector, even though, still poorly performs, the boosted performance from the model architecture serves as an indication to consider more sophisticated structures and signals. Therefore, the three Deep RL agents in Bons.ai with decent statistics show great potential to be turned into profitable trading strategies.

\subsection{Technical and Fundamental Features}

A good complement to the Bons.ai platform is the in-house code using a DQN agent. The in-house code provides a more granular level of control. To match the training speed in Bons.ai, we leverage High Performance Computing (HPC) with identical walk forward schemes. Additionally, a local version has also been implemented with an update frequency based on yearly retraining due to limited local compute resources. Furthermore, for the analysis of technical and fundamental features, the DQN agent and two traditional machine learning benchmarks are compared with only technical features and with technical and fundamental features. All models use the sliding-window training approach, and DQN moves forward every year (identical to the Bons.ai version), whereas Linear Regression and Random Forest models are optimised with a four-month sliding-forward window. The performance of the respective models is compared in Table \ref{fig:dqn_benchmark}. 

\begin{table*}[h]

\begin{center}\begin{tabular}{|c|c|c|c|c|c|}
\hline 
{\bf Agent} & {\bf Sliding Window} & {\bf Retrain Freq.} & {\bf Avg. SR.} & {\bf MDD (\%)} & {\bf Cum P\&L (M\textsterling)} \\\hline 
DQN (tech.) & 1 Year & Yearly & $0.96 \pm 0.24$ & $20 \pm 4$ & $24 \pm 4$ \\\hline 
DQN (tech.+ fund.) & 1 Year & Yearly & $0.94\pm0.41$ & $20\pm6$ & $24\pm7$ \\\hline
Lin.Reg (tech.) & 4 Months & Daily & $0.53\pm0.25$ & $24\pm4$ & $22\pm2$ \\\hline
Lin.Reg (tech.+ fund.) & 4 Months & Daily & $0.66\pm0.36$ & $21\pm5$ & $27\pm3$ \\\hline
R.F. (tech.) & 4 Months & Daily & $0.91\pm0.14$ & $19\pm3$ & $23\pm3$\\\hline
R.F. (tech.+ fund.) & 4 Months & Daily & $0.84\pm0.26$ & $25\pm5$ & $20\pm4$ \\\hline
 
\end{tabular} 

	\vspace{0.5pt}
	\caption[DQN vs Benchmark]{Result table for in-house DQN with traditional machine learning baseline models as benchmarks, based on technical features only (tech.) and with technical plus fundamental features (tech. + fund.). All models use the sliding-window training approach, and DQN has yearly sliding-window size (identical to the Bons.ai models), whereas Linear Regression (Lin.Reg) and Random Forest (R.F.) models have 4-month size. The average Sharpe Ratio, maximum drawdown and cumulative P\&L are reported, with figures after '$\pm$' sign denoting the respective standard deviation. Averages and standard deviation are taken over yearly samples from 2013 to 2020.}
	\label{fig:dqn_benchmark}
\end{center}
\end{table*}

The addition of fundamental features does not seem to improve strategy performance. For pure technical-feature-based agents, the DQN surpasses the Linear Regression agent, especially in terms of the average Sharpe Ratio (0.96 against 0.53). Yet, Random Forest seems to be comparable in all three metrics. The two traditional machine learning agents have a much faster daily retrain update frequency because of their low computational cost, and the result suggests Random Forest is a viable alternative to the DQN agent. 

\begin{figure}[h!]
	\centering
	\includegraphics[scale=0.25]{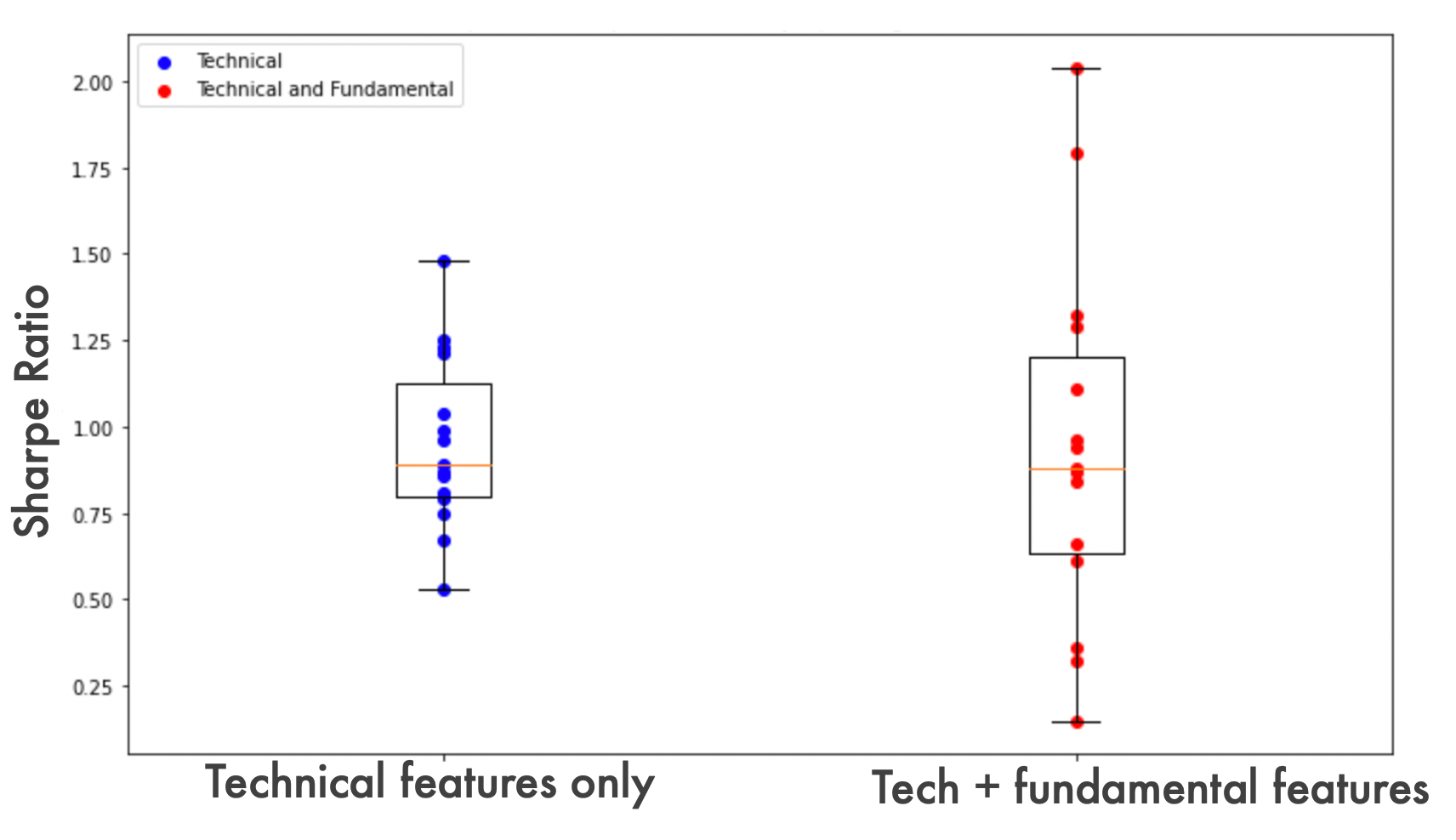}
	\vspace{0.5pt}
	\caption[DQN with technical vs. DQN with technical and fundamental features]{Sharpe Ratio box plot for DQN with technical features only and with technical plus fundamental features.}
	\label{fig:TechvsFund}
\end{figure}

To further discuss the role of fundamental features, a box-plot of Sharpe Ratios based on 15 realizations is presented in Fig.~\ref{fig:TechvsFund}, where one agent is based purely on price features with technical indicators, and the other agent has in addition also fundamental features.

Judging from the comparison of average Sharpe Ratios in Table \ref{fig:dqn_benchmark} and the median Sharpe Ratio indicated in Fig.~\ref{fig:TechvsFund} by the red line in the middle of each box, the difference in Sharpe Ratio with and without fundamental features is subtle. However, including fundamental features results in almost double the standard deviation when compared to the results with technical features only. One possible explanation could be that more features increase the agents' search space, and hence increase the variance of the performance.

\subsection{Ensemble Learning} \label{sec: REL}

\begin{figure}[h!]
	\centering
	\includegraphics[scale=0.3]{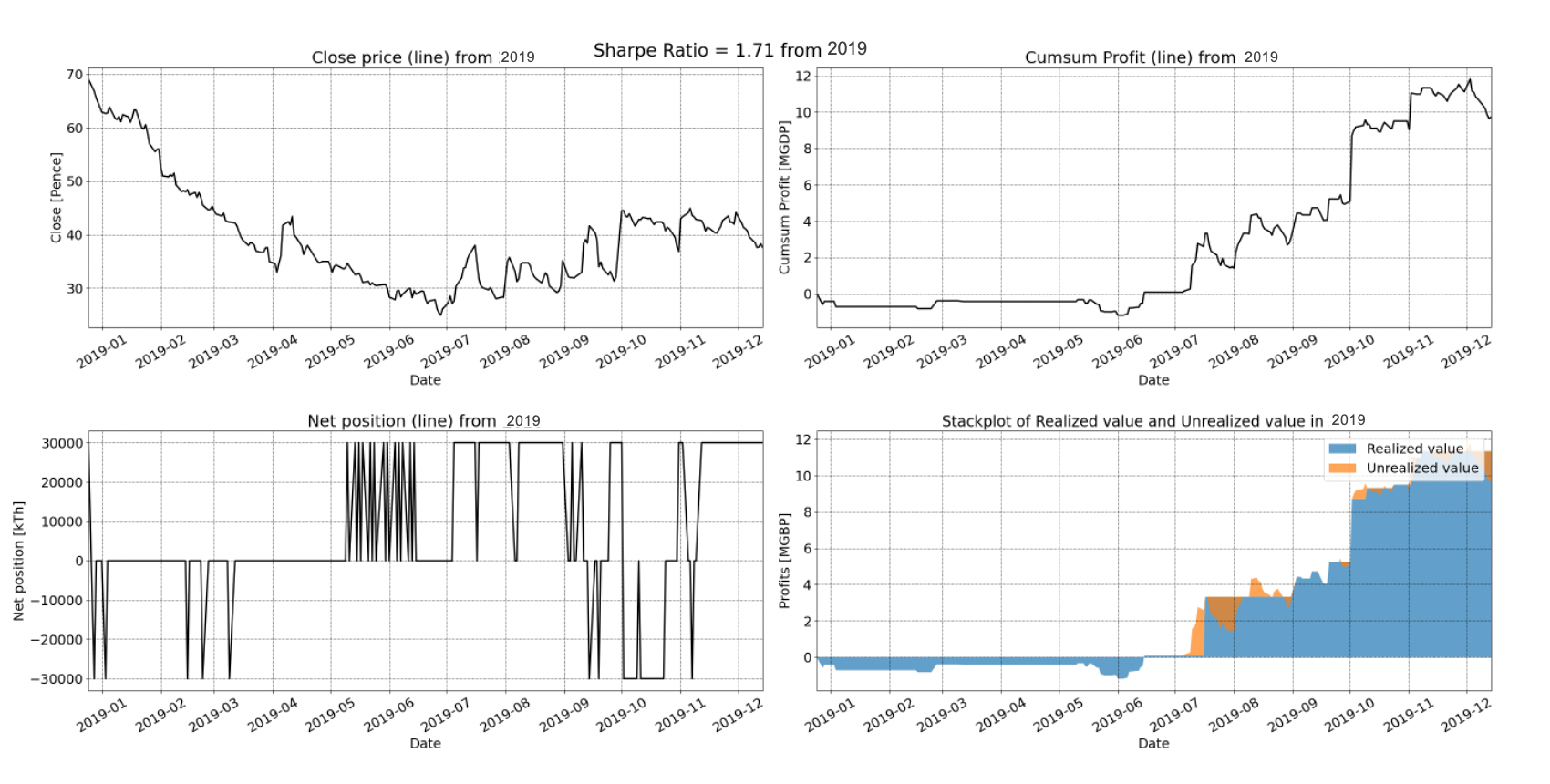}
	\vspace{0.5pt}
	\caption[Ensemble Learning Result 2019]{Ensemble Learning Results 2019 contains four blocks, that are the daily close price (top left), cumulative return (top right), net position for holdings (bottom left), and realized and unrealized returns (bottom right).}\label{ELR19}
	
\end{figure}

Each ensemble trains ten instances of the Deep RL agent overall episodes of the two years of training data preceding the test data set. Three ensemble learning results in 2018, 2019 and 2020 are performed. Each ensemble starts with ten realizations before filtering, and only 1, 0, and 2 agents are filtered out due to inferior training performance in 2018, 2019 and 2020, respectively. Back-testing performance of 2019 is presented in Figure \ref{ELR19} for exemplification.

Illustrated in Figure \ref{ELR19}, training in 2017 and 2018 produces a back-testing result with a Sharpe Ratio of 1.71 in 2019. For conciseness, we defer the results for 2018 with a Sharpe Ratio of 1.39 and for 2020 with a Sharpe Ratio of 0.49 to Appendix \ref{AppA: ELR18} and Appendix \ref{AppB: ELR20}, respectively. Moreover, in all cases, a positive P\&L is achieved. From the top left subplot, it is obvious that the market in the first half of 2019 seriously plummeted, while it oscillated in the second half of 2019. Under this scenario, the ensemble agent first attempted to go long which resulted in negative returns. Then, it intends to follow the trend by taking consecutive small holding periods of shorts in the first half. However, none of these attempts seemed to be effective enough to bring a positive P\&L. In the second half of 2019 before October, the agents performed multiple good trades by going long, which has seized most of the opportunities when the oscillations peaked. Then, both long and short trades have consolidated their gains at the end of 2019.

This analysis has reviewed certain characteristics of our ensemble trading bot in 2019. Most of its winning trades are based on mean-reversion behavior in the second half, and it failed trend-following in the first half. However, it does not seem trivial to infer a rule-based strategy. 
%Nevertheless, based on the empirical results from 2018 and 2020 it does not seem trivial to infer a rule-based strategy. 
Yet, as suggested by Wang \citep{Wang2021DynamicPO} and showcased in Section \ref{sec:ME}, the market exhibits different behaviours in different periods, and our agents should learn which features to rely on and which trading patterns to follow dynamically.

\section{Model Explainability} \label{sec:ME}
\begin{figure}[h]
        \centering
        \includegraphics[scale=0.2]{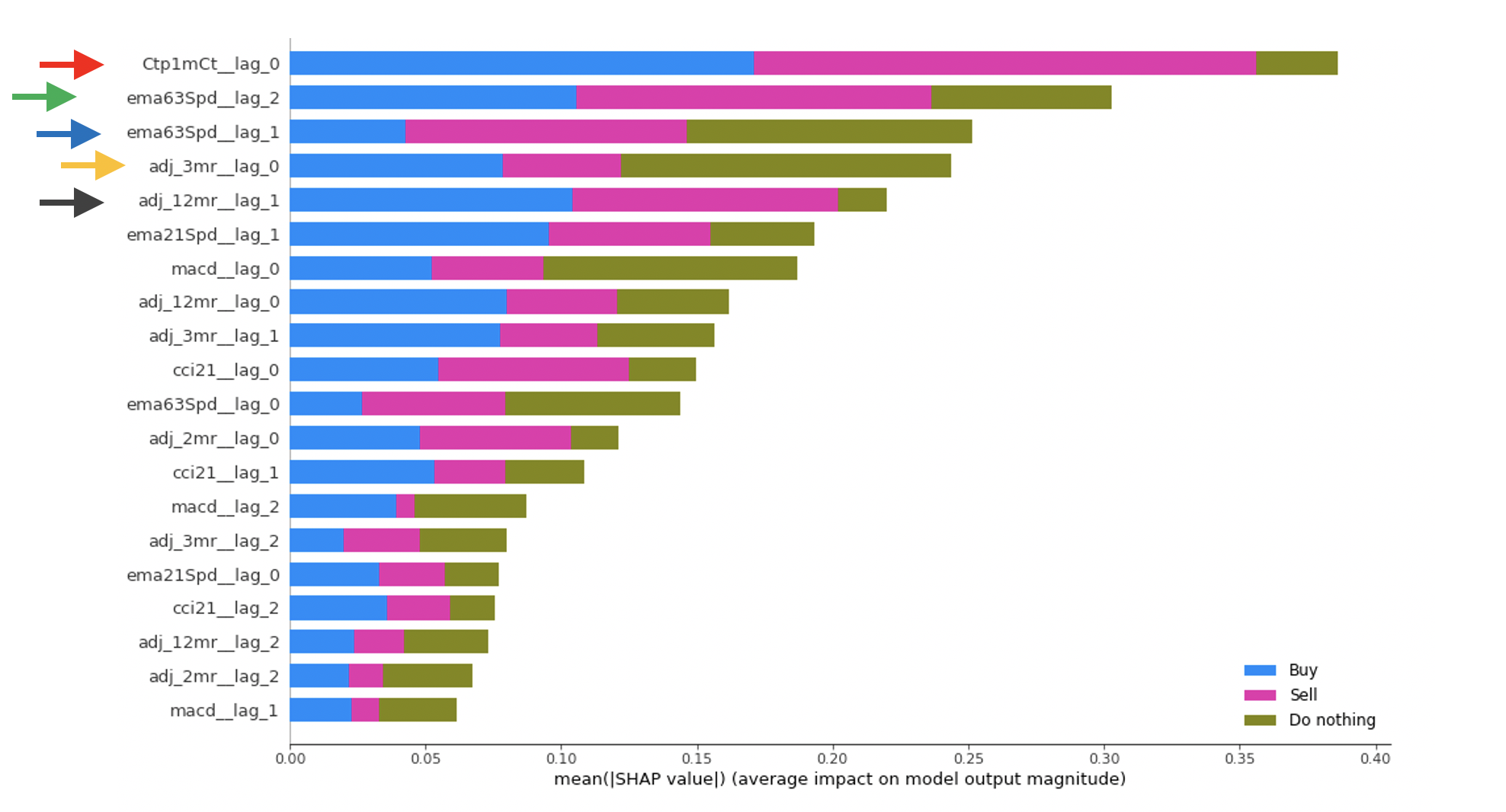}
        
        \vspace{0.5pt}
	\centerline{2014}
        \centering
        \includegraphics[scale=0.2]{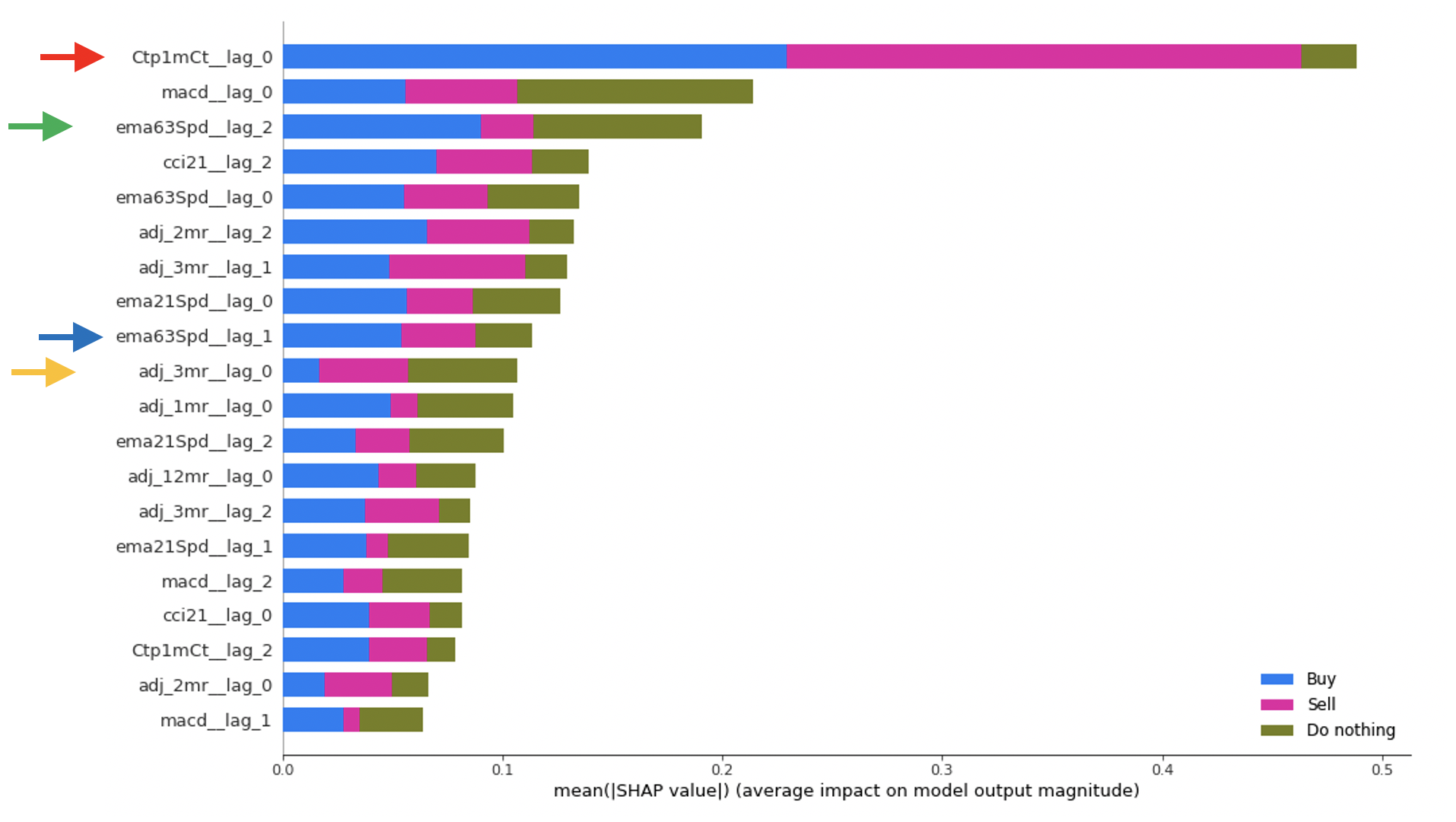}
        
        \vspace{0.5pt}
	\centerline{2020}
        \caption[Shapley2020]{Shapley feature contribution plot for 2014 \textit{(top)} and 2020 \textit{(bottom)}. The blue, purple and green bar indicates the feature importance for Buy, Sell and Do nothing decisions. The highlighted and tracked features are relative close price between t and t-1 with no lag (red arrow), spread between close and 63-day EMA (exponential moving average) with 2-day lag (green arrow), spread between close and 63-day EMA with 1-day lag (blue arrow), 3-month volatility-adjusted return with no lag (yellow arrow) and 12-month volatility-adjusted return with 1-day lag (yellow arrow).}
         \label{fig:shapley 2014 2020}
\end{figure}

To analyse feature importance and their contribution at various points in time, SHAP plots are used for ad-hoc model explainability analysis. In Fig.~\ref{fig:shapley 2014 2020}, the relative feature importance of 2014 and 2020 is presented. 

We analyze the relative feature importance of 2014 and 2020. The top 5 features in 2014 are relative close price between t and t-1 with no lag (feature 1), spread between close and 63-day EMA (exponential moving average) with 2-day lag (feature 2), spread between close and 63-day EMA with 1-day lag (feature 3), 3-month volatility-adjusted return with no lag (feature 4) and 12-month volatility-adjusted return with 1-day lag (feature 5).

Feature 1 remains in the dominant position in both years, with a marginal increased influence in Buy. However, features 2, 3, 4 become less significant in 2020 compared 2014, while the feature 5 has dropped out of the top 20 features. This progression suggests that feature importance is not constant across the time-series, and regular re-training is necessary for models to reflect most recent market information. Moreover, it is worth noting that except for the feature 1 which does not seem to change its mean SHAP value, the SHAP values of the other features are all smaller in 2020 than they were in 2014. This distribution may indicate that certain features dominate or are highly influential in 2014, while they tend to contribute equally in 2020. As a result, we would expect a better performance in feature selection in 2014 than in 2020 since fewer features are required to approximate the full model.

Furthermore, besides the feature contribution analysis in 2014 and 2020, a temporal decision plot is also visualized and used for our analysis. The full visualization is in the format of a video and provides details on how much a given feature contributes to a potential Buy, Sell or Hold decision. Not only can it serve as a verification for the feature selection process, but it can also be used as a tool to explain the model's rationale to stakeholders and non-technical parties.

\section{Discussion} \label{sec:Discussion}

\begin{figure}[h!]
	\centering
	\includegraphics[scale=0.27]{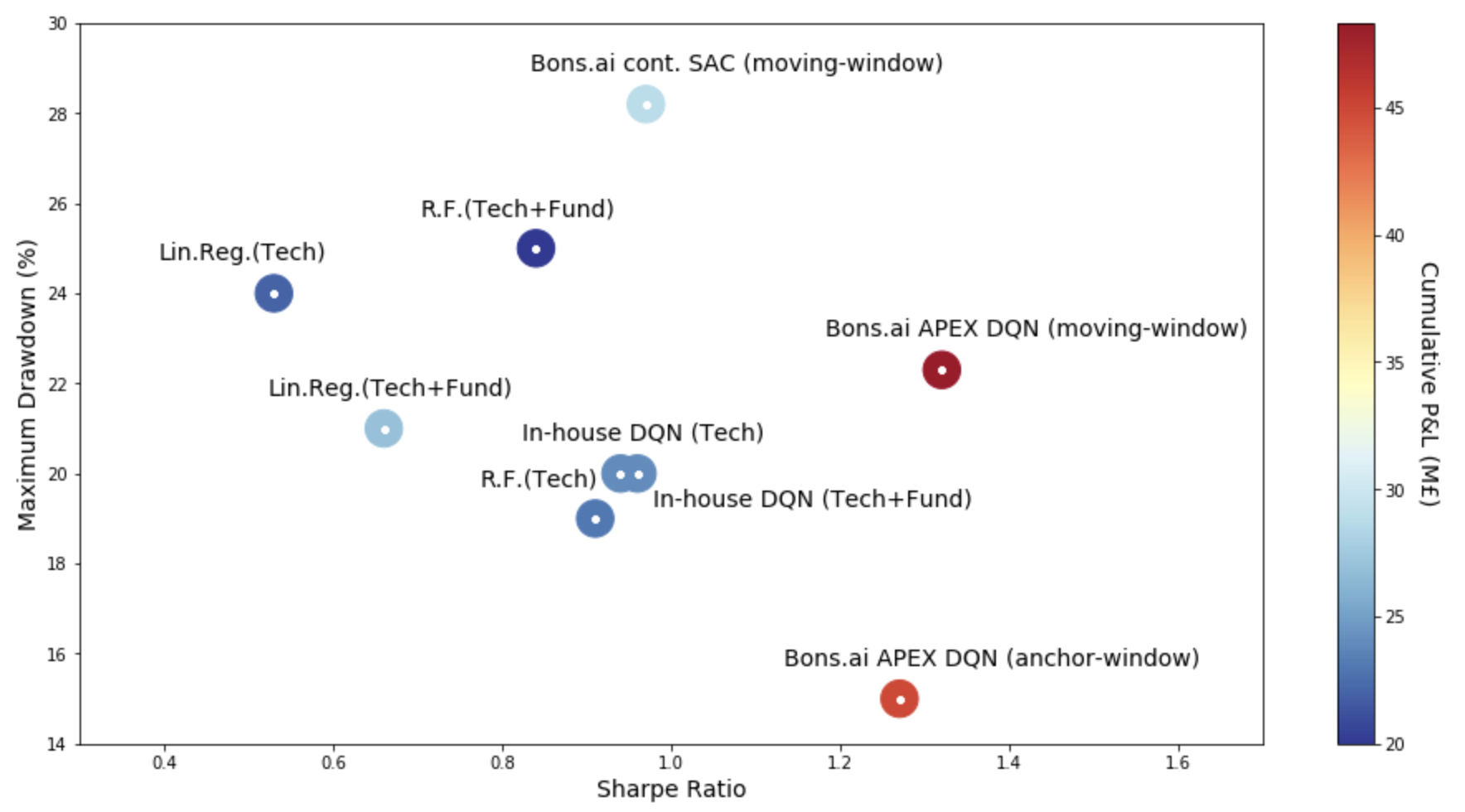}
	\vspace{0.5pt}
	\caption[Agent Summary Plot]{The summary plot for all machine learning trading agents including RL agents and traditional ML agents assessed over the period from 2013 to 2020. The x-axis represents the Sharpe Ratio, y-axis represents the maximum draw-down, and the color represents the cumulative P\&L. Lin.Reg represents linear regression agent, R.F. represents random forest agent, Tech indicates only technical features are used, Tech+Fund indicates technical and fundamental features are used in the in-house agents. All Bons.ai agents use technical and fundamental features.}\label{fig:AgentSumPlot}
	
\end{figure}

The performance of the Deep RL agents, the Linear Regression and the Random Forest models are compared in Fig.~\ref{fig:AgentSumPlot}. The x-axis represents the Sharpe Ratio, y-axis represents the maximum drawdown, and the color represents the cumulative P\&L. The best result with a Sharpe Ratio of $1.32$ and a maximum drawdown of $22.3\%$ is achieved by Bons.ai DQN Apex with moving window for re-training. The other results obtained generally also demonstrate decent Sharpe Ratios. We find an average Sharpe Ratio of $1.07$ taken across all versions and algorithms, which outperforms the state-of-the-art result in Zhang  \citep{Zhang2019DeepRL} where a Sharpe Ratio of $0.723$ for commodity trading with DQN has been reported. Fig.~\ref{fig:AgentSumPlot} also displays more conventional ML strategies based on Linear Regression and Random Forest. However, the Deep RL based strategies appear to be superior for this gas trading use case. Moreover, by means of ensemble learning in the form of Filtered-Thresholding we can further improve the performance. Backtesting yields an average Sharpe Ratio of $1.20$ over 2018-2020, a $23\%$ increase from the average Sharpe Ratio of $0.975$ obtained with the in-house DQN method. 

Despite the results obtained outperform those reported in state-of-the-art literature, implementing a Deep RL based trading agent faces several challenges in practice. First, there is usually still a drop in performance when going from back-testing to live trading. Second, extended periods of underwater performance would call for shutting down an algorithm before it can swing to profitability. In other words, even if in backtesting it generated a profit over the entire year of 2019, see Fig.~\ref{ELR19}, in practice it would not even reach that point, since it would have been stopped out well before. Third, the frequency of trades has to fit with the overall strategy of the trading desk and neither display overly long holding periods nor too frequent trades / churn. Fourth, model explainability remains a concern for black-box neural network based models although the analysis based on SHAP values in Section~\ref{sec:ME} and feature importance help to mitigate this point.

\section{Conclusions}
\label{sec:Conclusions}

The topic of systematic trading of commodities presents a challenge in quantitative trading due to the low signal-to-noise ratio, which makes learning models susceptible to overfitting. In this study, we present our implementation of a Deep Reinforcement Learning framework for systematic gas trading using different approaches based on the Microsoft Bons.ai platform and in-house code. Our Deep RL agent trained in Bons.ai has achieved a Sharpe Ratio of $1.32$ in back-testing, surpassing state-of-the-art results in the literature. We also proposed an ensemble learning scheme for our in-house DQN method, which achieved a Sharpe Ratio of $1.2$ and a $23\%$ improvement in performance over individual DQN agents.

Comparing models using only technical features with those using both technical and fundamental features, we found that including fundamental features did not lead to better performance. This may be due to the increased noise in the observation space caused by the less frequent and error-prone fundamental features. Another comparison aimed to discover different work-forward training schemes, and our results suggest the advantage of using a sliding-window approach. Despite the use of an experience replay buffer with Gaussian noise, we found that the amount of information extracted over noise introduced by a longer period of history decreases. Possible improvements could be made through a prioritized replay buffer or exponential moving average on features that emphasize the most recent knowledge.

This paper is one of the first to apply model explainability using Shapley values for Deep Reinforcement Learning in trading. It provides insight into which feature drives the agent's buy, sell or hold decision at a particular point in time, offering a way to analyze the developed rationale of an agent's trading strategy. The visualization confirms that technical features dominate the learning and inference in this specific domain problem.

Despite performance beyond state-of-the-art literature, implementing such Deep RL trading agent in practice presents several challenges, such as potential drops in performance between back-testing and live trading, extended periods of even slight under-performance triggering a shut down of the algorithm before it can reach profitability, the trading frequency having to fit with the desk's overall strategy, and model explainability. 

The successful application of Deep Reinforcement Learning to systematic gas trading highlights that rigorous feature selection, design of the reward function, model architecture, and ensemble learning can result in improved and robust performance. Ongoing and future work will consider the application of Deep RL to auction-like European power markets \citep{Wang2023DeepRL} and process optimization \citep{DeepRLProcessOptReport2022}.
\newpage
%%
%% The acknowledgments section is defined using the "acks" environment
%% (and NOT an unnumbered section). This ensures the proper
%% identification of the section in the article metadata, and the
%% consistent spelling of the heading.
\begin{acks}
The authors would like to acknowledge the contributions of and collaboration with Boris Lastdrager (Shell), Tina Zhao (Shell), Tashi Erdmann (Shell) and Hossein Khadivi Heris (Microsoft Bons.ai).

The authors also would like to acknowledge Jeremy Vila (Shell), Franz Kiraly (Shell) and Boris Lastdrager (Shell) for reviewing the manuscript and providing helpful comments and feedback.

Moreover, all authors would like to acknowledge the funding of the Shell.ai Futures Programme which has enabled this research and the Computational Sciences and Digital Innovations (CSDI) Lab for providing the compute resources (Azure).
\end{acks}
\newpage
%%
%% The next two lines define the bibliography style to be used, and
%% the bibliography file.
\bibliographystyle{ACM-Reference-Format}
\bibliography{sample-base}
\newpage
%%
%% If your work has an appendix, this is the place to put it.
\appendix
\section*{Appendixes}
\renewcommand{\thesubsection}{\Alph{subsection}}
\subsection{Appendix A: Ensemble Learning Result 2018} \label{AppA: ELR18}
\begin{figure}[h]
    \centering
    \includegraphics[scale=0.3]{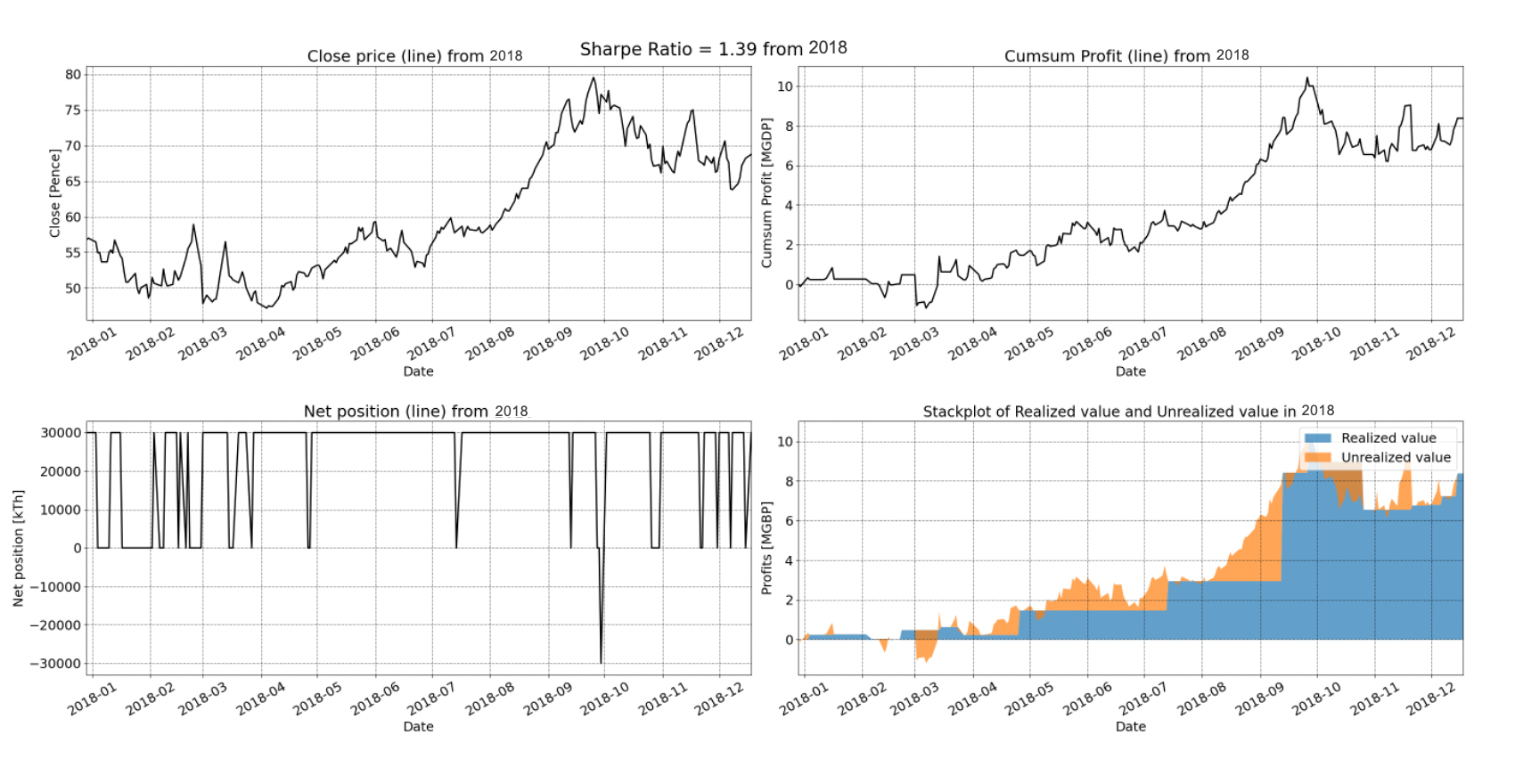}
    \vspace{0.5pt}
    \caption[Ensemble Learning Result 2018]{Ensemble Learning Results 2018 contains four blocks, that are the daily close price on the top left, cumulative return on the top right, net position for holdings on the bottom left, and realized and unrealized returns on the bottom right.}\label{ELR18}
        
\end{figure}
\subsection{Appendix B: Ensemble Learning Result 2020} \label{AppB: ELR20}
\begin{figure}[h]
    \centering
    \includegraphics[scale=0.3]{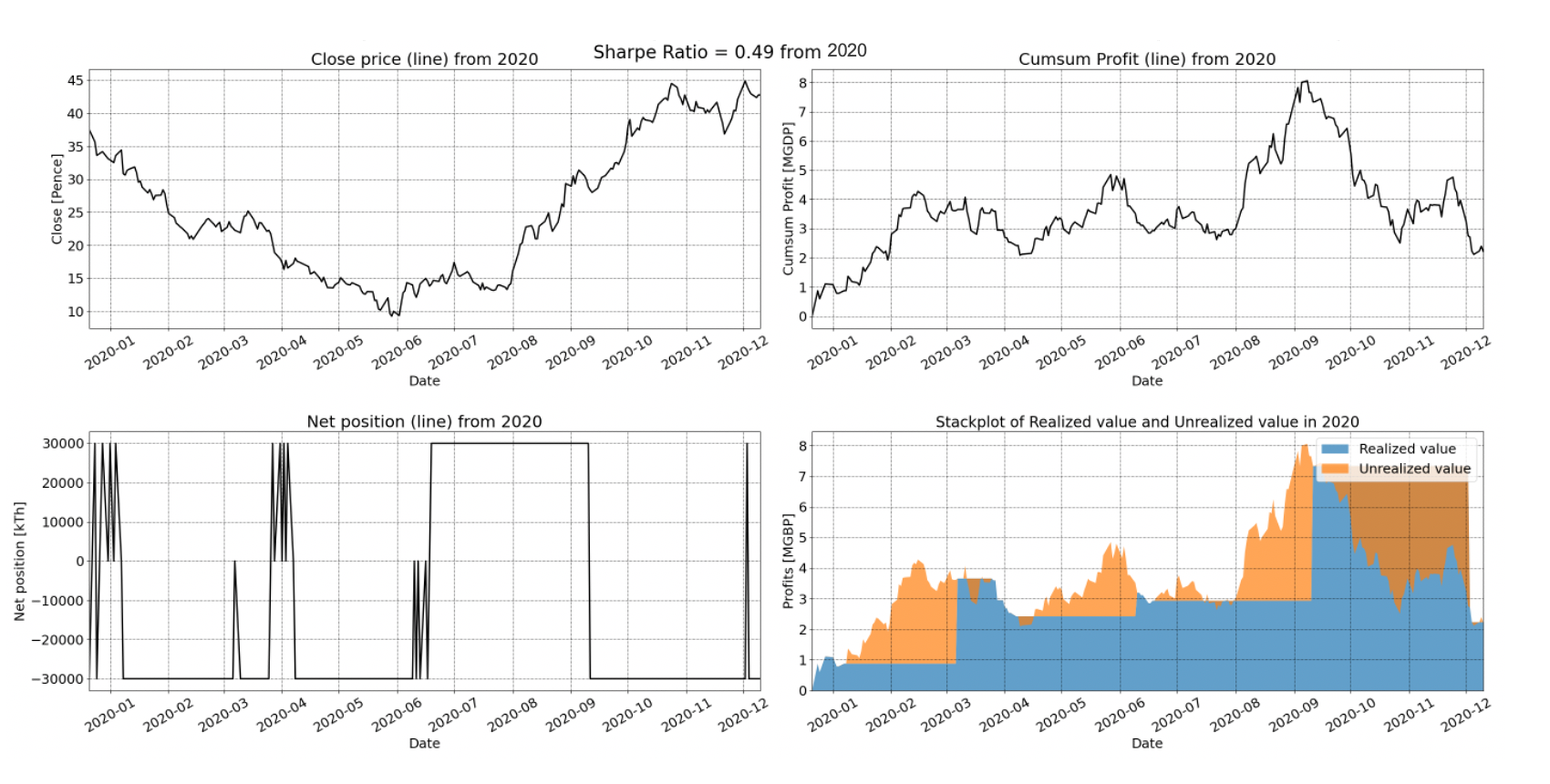}
    \vspace{0.5pt}
    \caption[Ensemble Learning Result 2020]{Ensemble Learning Results 2020 contains four blocks, that are the daily close price on the top left, cumulative return on the top right, net position for holdings on the bottom left, and realized and unrealized returns on the bottom right.}\label{ELR20}
        
\end{figure}

\end{document}